\documentstyle[sprocl]{article}
\def\Journal#1#2#3#4{{#1} {\bf #2}, #3 (#4)}

\def\PLB{{\em Phys. Lett.}  B}

\def\PRD{{\em Phys. Rev.} D}

\def\be{\begin{equation}}
\def\ee{\end{equation}}
\def\bea{\begin{eqnarray}}
\def\eea{\end{eqnarray}}

\begin{document}
\title{STRING THEORY DUALITIES}
\author{ Michael Dine }
\address{Santa Cruz Institute for Particle Physics \\
University of California \\ Santa Cruz, CA   95064}
\maketitle\abstracts{
The past year has seen enormous progress
in string theory.  It has become clear that all of the
different string theories are different limits of a single
theory.   Moreover, in certain limits, one obtains
a new, eleven-dimensional structure known as $M$-theory.
Strings with unusual boundary conditions, known as
D-branes, turn out to be soliton solutions of string theory.
These have provided a powerful tool to probe the structure
of these theories.  Most dramatically, they have yielded
a partial understanding of the thermodynamics of black
holes in a consistent quantum mechanical framework.
In this brief talk, I attempt to give some flavor of these
developments. }

\section{Introduction}

The past two years have been an extraordinary period for those
working on string theory.  Under the rubric of duality, we have
acquired many new insights into the theory.\cite{apology}
Many cherished
assumptions have proven incorrect.  Seemingly important
principles have turned out to be technical niceties, and structures
long believed irrelevant have turned out to play a pivotal
role.  Many extraordinary connections have been discovered,
and new mysteries have arisen.  For phenomenology,
we have learned that weakly coupled strings may be
a very poor approximation to the real world, and that a better
approximation may be provided by an $11$-dimensional
theory called $M$ theory.  But perhaps
among the most exciting developments,
string theory for the first time has
begun to yield fundamental insights about
quantum gravity.

Before reviewing these important developments, it
is worthwhile recalling why string theory is of interest
in the first place.  There are parallels in our
current situation vis a vis string theory
and the situation $30$ years ago in the theory of weak
interactions.\cite{apologya}
For many years, it was clear that the four fermi interaction
was not renormalizable, and that this was the signal
of some new physics at energy scales below a $TeV$.
Moreover, it was widely believed that this new physics involved
the exchange of a vector boson.   Gradually, it became clear
that only in theories with non-abelian gauge bosons
are such exchanges consistent.  The requirement
of consistency, in other words, strongly
constrained the nature of the microscopic theory.
It only was necessary to determine the gauge group,
and this could be done by studying experimentally
and theoretically the interactions of the light
states of the theory at low energies.  Of course, until
the discovery of the $W$ and $Z$, one might
have wondered whether we had missed something,
whether there might be some other structure
which permitted massive vector fields, but in the end,
this was not the case.

The current situation with respect to gravitational interactions
is similar.  General relativity cannot be a consistent
theory up to arbitrarily high energies.  We know only
one microscopic structure which can consistently incorporate
gravity and gauge interactions:  string theory.  We do
not know for certain that there are not other structures,
but -- as we will in some sense see in this talk -- it is
quite possible there are not.  Of course, it will be a long
time before we can do direct experiments on Planck scale
physics, but we might hope to elucidate many features
of the microscopic theory by studying its low energy phenomenology.

It is remarkable that postulating that the fundamental entities
in nature are strings rather than point particles
{\it automatically} gives
\begin{itemize}
\item  Gravity.
\item  Gauge interactions, with a gauge group ${\cal G}$
which is large enough to contain the $SU(3) \times SU(2) \times
U(1)$ group of the standard model.
\item   Finite theories with good high energy behavior.
\end{itemize}

More evidence that the theory might be true comes from
studying classical solutions.  Particular solutions give:
\begin{itemize}
\item  Repetitive generations (sometimes 3).
\item  Light Higgs:   An essential piece of hierarchy
problem.
\item  Axions:  The theory automatically has Peccei Quinn
symmetries which hold to a sufficiently good approximation
to solve the strong CP problem.
\item  N=1 Supersymmetry:  String theory doesn't make
sense without supersymmetry.  In fact, we don't really know
how to make sense of the theory unless supersymmetry
survives to low energy.  So it is probably fair to call
low energy supersymmetry a {\it prediction} of string theory.
\end{itemize}

These are extraordinary achievements for  a theory, and
make it quite plausible that string theory might be
some sort of ultimate description of nature.  There are,
however, serious difficulties which must be overcome before
the case can be convincing:
\begin{itemize}
\item  We don't really know what the theory is.  It is as
if we had Feynman rules for a theory like QCD, but don't
know the field theory.  There are some questions we can
address, but many more which we cannot.  The recent
developments are providing new insights, but
we still seem a long way from a complete answer.
\item  Too many vacua.  Among these, there are a large
number of discrete choices as well as continuous choices.
Recent developments have given us some insight into
the meaning of the discrete choices.
\item  At weak coupling, which is the only regime in which
one can make real calculations, one inevitably makes
some predictions which are qualitatively wrong.
In particular, one predicts that the vacuum is unstable.
\item  Unification of couplings:  It is well known that
low energy supersymmetry leads to successful unification
of couplings, with a unification scale of order $2-3 \times 10^{16}
\rm GeV$.  In weakly coupled string theory, one expects
unification, but at a
scale of order $30$ times higher.  The argument is quite
simple.  The dimensionless coupling constant of string theory
is related to the unified coupling, $\alpha_{GUT}$,
the compactification radius, $R \approx M_{GUT}$,
and the tension, $T$
by
\begin{equation}
\alpha_{st}= \alpha_{GUT} R^6T^3.
\end{equation}
If we require $\alpha_{st}<1$, then
$R^2 \approx T^{-1} \approx 6 \times 10^{17} \rm
GeV$.\cite{dienes} We will see that the recent developments in string duality
suggest a solution to this problem.
\end{itemize}

Almost all of these points could have been made years
ago.\cite{dinedpf}
In the last two years, however, there have been striking
developments which bear on each of the points 1-4,
associated with ``Duality."  Duality is a term which is
used, loosely, to define equivalences between different
physical theories.  The recent explosion of activity involves
many kinds of dualities:  equivalence between different
string theories, between theories with different values of
couplings, compactification radii, and, perhaps most
surprisingly, equivalences between string theories and
certain -- as yet poorly understood -- theories in eleven
dimensions.  If there has been a general
theme underlying these efforts, it has been to exploit
the huge degeneracy of vacua and the symmetries of the
theory(ies) to gain insight.

This work has already taught us {\it many} lessons, but
there are two which are particularly striking:
\begin{itemize}
\item  There is only one string theory; all of the
previously known theories are equivalent.  The fact that
all of the theories of gravity we previously knew are equivalent
suggests that there is only one consistent theory
of gravity!
\item  If ``string theory" does describe the real world,
weakly coupled strings are likely to be a very poor approximation.
A better description can be obtained
in terms of an $11$-dimensional theory, only
some of whose features we know.  This theory is referred
to as ``M-theory."
\end{itemize}

These developments are also finding an application:
for the first time, we are making a controlled attack
on one of the fundamental
problems of general relativity:  the thermodynamics of
black holes and the problem of information loss.
One aspect of this problem is that black holes behave as
if they possess an entropy, the
famed Beckenstein-Hawking entropy:
\begin{equation}
S={A \over 4 G}.
\end{equation}
Up to now, the significance of this entropy has been obscure.
However, for certain black holes in string theory, it has been
possible to count the degenerate ground states, and
show that one obtains eqn. 2.

In the rest of this talk, I will give a brief overview of these
developments.  Progress has been extraordinary; one
could easily write several books on the subject.
In $30$ minutes, I must, of course, be highly selective.
Indeed, the usual apology that one can't cover everything
is more heartfelt here than usual.  For string duality, at the
moment, represents a large collection of beautiful observations,
but the big lesson they are teaching us is not entirely clear.
It is quite possible that the most important lessons may
lie in things which I omit.  That said, I will first briefly
remind you about electric-magnetic duality.  I say ``remind"
because this is a topic you can find covered nicely
in Jackson.\cite{jackson}
I will then turn to the interconnection of different
string theories.  I will discuss the large degeneracy of string
vacua (associated with ``moduli"), and describe how,
as one moves around this ``moduli space," one
encounters all of the different string theories.  I will
then explain the connection between ten and eleven
dimensions, and discuss some phenomenology of $M$ theory.
Towards the end, I will discuss a new tool for studying
non-perturbative questions in string theory:  ``D(irichlet)-branes,"
and the application of this tool to the black hole information
problem.  I will conclude with a listing of some recent
developments, and deep questions.

\section{Electric-Magnetic Duality}

We have all stared at Maxwell's equations and wondered
whether there might be magnetic charges.  In the presence
of magnetic charges and currents, one has\cite{jackson}
\begin{equation}
\vec \nabla \cdot \vec E = \rho_e~~~~~
\vec \nabla \cdot \vec B = \rho_m
\end{equation}
\begin{equation}
\vec \nabla \times \vec B
- {\partial \vec E \over \partial t} = \vec J_e~~~~~
\vec \nabla \times \vec E
+ {\partial \vec B \over \partial t} = -\vec J_m.
\end{equation}
These equations possess a symmetry, under the
replacements
\begin{equation}
\vec E \rightarrow \vec B
~~~~~~\vec B \rightarrow -\vec E
~~~~~\rho_e \rightarrow \rho_m ~~~~~
\vec J_e \rightarrow \vec J_m ~~~~~\vec J_m \rightarrow
- \vec J_e.
\end{equation}

What sort of symmetry is this?  In field theory, monopoles
arise as solitons.  Their masses behave as
\begin{equation}
M_m= {1 \over e^2} v.
\end{equation}
They
are big fat objects, which obey the Dirac quantization
condition, $eg=2 \pi n$, where $g$ is the magnetic charge.
If duality is to be a symmetry, it must somehow interchange
``fundamental" particles and solitons.  This is not completely
crazy, since it also must interchange $e \leftrightarrow {2 \pi}/e$,
i.e. electric-magnetic duality is weak-strong coupling duality.
On the other hand, because of this, it is hard to see how duality can be
more than a speculation.

It turns out, however, that in theories with enough supersymmetry,
one can check the duality conjecture.In such theories, the supersymmetry algebra takes the form,
\begin{equation}
\{Q_{\alpha}^I, Q_{\beta}^J\}
=P_{\mu} \gamma^{\mu}_{\alpha \beta}\delta^{IJ} + \epsilon_{\alpha \beta}
Z^{IJ}~~~~~I,J=1,\dots ,N.
\end{equation}
Here the $Z$'s are some set of charges (e.g. the electric
and magnetic charges) which are referred to as ``central
charges."  If one has a soliton which is invariant under some
of the $Q$'s, one can prove exact formulas for the mass,
called BPS formulas.
The basic point is quite simple.
Schematically,
\begin{equation}
\langle \{ Q_{\alpha}Q_{\beta}\} \rangle = 0
= \langle H \rangle + q.
\end{equation}
$\langle H \rangle$ is just the mass and $q$ is
a charge, so the mass
is related to a charge.  Now one can determine
the mass at weak coupling, interpolate to strong
coupling, and verify the duality conjecture.  This type of analysis,
first performed in field theory, can be extended to string
theory.

So there are theories in which electric-magnetic duality
holds.  Solitons are mapped to ``fundamental" particles
under these transformations.  The duality symmetry
can be thought of as a spontaneously broken symmetry.
It is restored if $e = \sqrt{2 \pi}.$

\section{Moduli Spaces and String Equivalences}

In string theory, there are a variety of dynamical fields,
referred to as moduli,
whose expectation values determine the parameters
of the theory.  They have the property that in some
lowest order approximation, they have no potential;
in many cases (particularly if there is a high
degree of supersymmetry), one can argue that
they have no potential exactly, i.e. even when all non-perturbative
effects are taken into account.  This phenomenon
is not familiar in conventional, non-supersymmetric
field theories, so it is perhaps best to illustrate with some examples:
\begin{itemize}
\item  The dilaton.  In string theory, there is a field,
usually denoted by $\phi$, called the dilaton.  The expectation
value of this field determines the dimensionless coupling
of the theory, through an equation of the form
\begin{equation}
\langle e^{\phi} \rangle = g_s.
\end{equation}
\item  When one compactifies string theories, the
size and shape of the compact spaces are determined
by additional moduli fields.
\end{itemize}

The second phenomenon is nicely illustrated by
compactification of a $10$ dimensional string
on a circle of radius $R$.  From the perspective
of a nine-dimensional physicist, there are a number of
massless states. For example, for the components
of the ten-dimensional metric one has the
decomposition:
\begin{equation}
g_{MN}(x,\theta) \rightarrow g_{\mu \nu}(x)~~~~~g_{\mu 9}(x)
=A_{\mu}(x)~~~~~g_{99}=R^2(x).
\end{equation}
The radius, $R$, can be thought of as the expectation value
of the field, $R^2(x)$, i.e. the radius is dynamical.
Also it can take any value, so there is no potential
for this field.

There are dualities associated with $R$.  These dualities
are easy to establish, since (unlike
weak-strong duality) they are already visible in perturbation
theory.  If we compactify on a circle, we have momenta,
$p^9= {n/R}$.  From the perspective of a nine dimensional
observer, a {\it ten} dimensional massless field (such as the
metric) with momentum $p_9$ has mass $p_9^2$.
There are also windings, corresponding to the
fact that the string can wind $m$ times around the circle.
The mass spectrum is given by
\begin{equation}
M^2 = T ({n^2\over R^2} + m^2 R^2).
\end{equation}
This spectrum is symmetric under $R \rightarrow 1/R$.
Compactifying more dimensions yields more elaborate
dualities; these are generically referred to as $T$-dualities.
(Strong-weak coupling dualities are called $S$-dualities;
duality transformations which mix coupling and moduli
are called $U$-dualities).

With these preliminaries, we are in a position to discuss
the equivalence of the various string theories.  In textbooks,
one learns that there are five types of string theory:
heterotic $E_8 \times E_8$, heterotic O(32) , Type IIA and IIB (all
theories of oriented closed strings),
and the O(32) Type I theory (a theory of open and unoriented
closed strings).   For some time, it has been known that
the $E_8 \times E_8$ and heterotic O(32), as well as the
IIA and IIB theories, are related by $T$-dualities.
For example, compactifying the IIA theory on a circle
of radius $R$ gives the same theory as the IIB compactified
on a circle of radius $1 /R$.
More recently it has been realized
that the strong coupling limit of the O(32)
heterotic theory is the weakly coupled Type I theory.
This is rather amazing, since these theories are formulated
in terms of quite different objects (closed vs. open strings,
oriented vs. unoriented).  Similarly, suitable compactifications
of the $E_8 \times E_8$ theory at weak coupling are
equivalent to different compactifications of the type II
theories.  But perhaps most surprising of all, is that the
ten-dimensional Type IIA and the heterotic $E_8 \times E_8$
theories become, in the strong coupling limit, eleven
dimensional!

There is not time here to explore all of these connections,
but I would like to describe some of the features of the
10-11 dimensional duality.  The point, again, has to
do with the solitons of the theory.  In the IIA theory,
at weak coupling, there is a tower of solitons with
mass
\begin{equation}
M = {n \over g}
\end{equation}
for integer $n$.  These states
are BPS states, so the mass formula
is exact and holds even as
$g \rightarrow \infty.$
Equation $12$ is similar to the formula
for the momentum states
in Kaluza-Klein compactification, with $M= n/R$.
So $g \rightarrow \infty$ is similar to $R \rightarrow
\infty$.  So the IIA theory
resembles  an $11$ dimensional theory compactified
on a circle.  By more careful study, one can show
that this eleven dimensional theory is $11$-dimensional
supergravity (the only supersymmetric theory
in $11$ dimensions).  Note that now the dilaton and the
radius which we described before are more or less on the
same footing.  It turns out that the
radius of the eleventh dimension, $R_{11}$,
goes roughly as $R_{11} \sim g^{3}$.
In fact, this connection has been exploited
to provide a deeper understanding of various duality
symmetries.

For the heterotic string, the story is more intricate.
Again, one finds that the large coupling limit of the theory is
an eleven dimensional theory.  However,  the relevant
eleven dimensional world now has two walls, separated
by a distance $R_{11}$.  The graviton, metric and
other fields of eleven dimensional supergravity now
propagate throughout the eleven dimensional space,
but the gauge fields and gauginos live on the walls.
This picture is established by considering space-time
anomalies, the low energy spectrum and the
low energy effective action.

What is really going on
microscopically, from the eleven dimensional perspective,
is still not known.  Eleven dimensional supergravity is
not a renormalizable, much less finite theory, so it is
presumably the low energy limit of some other structure.
This structure has been called $M$-theory.
One might wonder, for example, whether the walls
described above are real walls, and the gauge fields
are states bound to them.

In any case, in this framework, we can solve the problem
of string unification.
Before we argued that
if $M_{GUT}$ was of order $10^{16}$ GeV, the
string coupling was enormous (of order $10^7$).  But
from our present perspective, this suggests that
we should consider the problem from the point of
view of $M$-theory.  The precise relations between
ten-dimensional and four dimensional quantities
are:
\begin{equation}
R_{11}^2 =  {\alpha_{GUT}^3 V \over 512 \pi^4 G_N^{2}
 },
\end{equation}
and
\begin{equation}
M_{11}= R^{-1} \left (2 (4 \pi)^{- 2/3} \alpha_{GUT}
\right )^{-1/6}.
\end{equation}
Plugging in reasonable
numbers gives that $M_{11} R \approx 2$,
while $M_{11} R_{11} \approx 70$.  Taking
these formulas at face value, the universe
is approximately five dimensional, and the
eleven dimensional supergravity approximation
should perhaps not be so bad!

 Even with this starting point, it is not so easy to develop
a detailed phenomenology, but there are at least two
immediate implications. 
\begin{itemize}
\item  The fundamental
scale of the theory is of order $M_{GUT}$, not $M_p$.
This raises issues for baryon number violation
and other effects mediated by high dimension operators.
\item  Axions:  The presence of axions is one
of the virtues of string theory.  However, most of the
Peccei-Quinn symmetries  are violated
by effects of order $e^{-R^2 T}$, and one usually
says that this is of order
one.  Now that we think of $R$ as large, this is not
the case, and these axions are viable.  They
have decay constants of order $M_p$, which presents
problems for conventional cosmology, but because of the presence
of the moduli, string cosmology is likely to be unconventional,
and it is not clear that these problems are so serious.
\end{itemize}

Eventually, we would like to understand why the scales
are what they are, i.e. what dynamics determines the
moduli expectation values.  There are hints of a possible
mechanism in recent work of Witten.  He showed that
if one holds the compactification volume and the coupling
of one of the gauge groups fixed, that of the other grows
with $R_{11}$, blowing up when
\begin{equation}
m_p = c {M_{GUT} \over
\alpha_{GUT}^{2/3}},
\end{equation}
with $c$ a number of order one.  This is not
an not unreasonable value.  Still, we are far from a complete
picture.

\section{D-Branes:  A New Tool}

One of the lessons of the recent developments is
that features of theories which appeared fundamental,
such as whether a theory contained closed or open
strings, are of no invariant significance.
Similarly, we have seen that solitons in one description
are fundamental entities in another.  Indeed, Polchinski
has observed that certain classical solutions of the string
equations can actually be described as fundamental
strings with unusual boundary conditions.\cite{polchinskietal}  

One can understand the appearance of $D$-branes by considering
open strings.  The usual free string action is
\begin{equation}
S={T \over 2} \int d^2 \sigma
\partial_{\alpha}X^{\mu} \partial_{\alpha}
X_{\mu} d^2 \sigma.
\end{equation}
There are two possible boundary conditions at the endpoints:
\begin{equation}
\partial_{\sigma} X^{\mu}=0
\end{equation}
\begin{equation}
X^{\mu}=Y^{\mu} (constant).
\end{equation}
The first of these are the usual Neumann conditions.  The
second, Dirichlet, condition, is usually discounted because
it violates translation invariance. But for the description
of solitons, this is fine.   The $Y^{\mu}$ correspond
to the locations of the soliton.  One can imagine that
the time and $p$ of the space components obey
Neumann conditions, while the remaining coordinates
obey Dirichlet conditions.  The resulting object is called
a Dirichlet $p$-brane (a $D0$ brane is a particle, a $D1$-brane
a string, a $D2$ brane a membrane, and so on).
What is striking
here is the simplicity of the $D$-brane description.
Complicated solitons -- and the quantum fluctuations
about them -- are described in terms of simple two
dimensional field theories.

This $D$-brane technology has found many applications.
Among the most interesting are to black hole physics.
There is a class of puzzles associated with the fact that black holes
behave, in many ways, as thermodynamic objects.
One can associate with them an entropy which obeys the usual
laws of thermodynamics.  They emit
particles like black bodies at that temperature.
This raises the possibility, however, that an initially
pure (albeit quite complicated) state which formed
a black hole might evolve into a thermal, mixed state.
The   puzzle is quite serious.  Simple explanations, such
as the possibility that the information is encoded in subtle
correlations among the emitted particles, run afoul
of principles of field theory such as locality and causality.

Various scenarios have been offered for how this problem might be resolved in
string theory, but they generally involve strong coupling,
and are difficult to discuss concretely.  With the recent
progress in string theory, however, some real steps
have been taken towards addressing these questions.
In several instances, the Beckenstein-Hawking
entropy, eqn. 2, has been shown to be equal to the
logarithm of the number of microstates.

The strategy in these calculations is not terribly
complicated.  One identifies soliton black holes
in the theory with certain configurations of $D$ branes.
The $D$ branes are described by
a free two dimensional field theory,
so the counting of degenerate ground
states is a reasonably straightforward problem.
One difficulty is that the calculation is only valid when the
coupling is weak, which turns out to correspond to
a Schwarschild radius much less than the string length.
In the interesting limit, the soliton picture is not valid.
However, we are rescued, again, by the fact that
the $D$-brane states are BPS states, so their masses
are correctly given, even at strong coupling, by
their weak coupling expressions.  Presently, many
workers are seeking to go beyond
the BPS limit in order to get a clearer
physical picture. 

The problems of black hole physics are important,
not so much in themselves, as for the challenge they
provide to our understanding of physics at very
short distances.  Hawking has long advocated the
view that they signal that quantum mechanics
itself must be modified in some drastic way.  In the
framework of conventional field theory, many workers
have tried, and failed, to address this challenge.
The fact that string theory is likely to resolve these
questions -- in a conventional quantum mechanical
framework -- is strong support for the idea that string
theory is the correct, underlying theory of gravity.

\section{Other Developments}

There have been a long list of additional beautiful
results, some of which may have profound significance.
Limitations of space (and of my knowledge) prevent
making any sort of complete list, but let me
mention a few (chosen largely because I want
to learn about them and understand them better):
\begin{itemize}
\item  There have been a number of applications
of duality to theories with $N=1$ supersymmetry
in four dimensions.   Dual pairs have been uncovered.
Mysterious phenomena in one picture (for example,
intricate cancellations) appear simple in another.
\item  New phases of theories have been uncovered.
Singularities in low energy effective
actions are usually associated with the appearance
of new light states.  Examples have been exhibited
where, for example, the topology of space-time
appears to change, and/or where monopoles or black
 holes become massless.  One also has examples where
an entire string-like tower of states becomes massless
(tensionless strings).
\item  Related to this, it has become clear that one can have
much larger gauge symmetries than are possible in weak coupling
strings. 
\item  One has obtained some insight into the meaning
of gauge symmetries.  Perhaps the most outstanding example
of this is due to Seiberg, who has exhibited field theories
with massless composite gauge bosons.
\item  Evidence for a new scale?  Perturbative
string amplitudes are very soft for momenta and distances
of order $\sqrt{T}$.  This is small compared to the
Planck scale, $M_p = \sqrt{T}/g.$  It is usually
said that it does not make sense to probe shorter distances,
and that the notion of space-time ceases to make sense
at this scale.  But duality
has obscured the significance of $T$, and there is evidence
for a harder, short distance component.
\end{itemize}

\section{Conclusions and Forecast}

Much has been learned in the last two years.
We know that there is only one string theory.  We have
greater insight into the moduli space of string vacua.
We have understood certain non-perturbative phenomena
in the theory.  String theories seem poised to meet
the challenge of black hole physics.  Yet we still
feel like the proverbial blind persons faced with
the elephant.  While we finally know that we are studying
one creature, not several, we still don't understand
quite what it is we have gotten a hold of.

Let me close, in the spirit of the times, with
a forecast.  Over the next year, I look for:
\begin{itemize}
\item  Further beautiful verification of string dualities.
\item  Persuasive resolution of the
black hole information loss problem.
\item  Perhaps something even more spectacular,
yielding greater insight into what exactly these
theories are.
\end{itemize}

But there are some questions which I am less optimistic
we will answer very soon.  The progress in duality involves
reformulating interesting strong coupling problems
as weak coupling problems.  Unfortunately, general
arguments suggest that if string theory describes nature,
{\it no} weak coupling analysis can be valid.
So I don't expect to see, for example, a calculation
of $m_e/m_{\mu}$ in the coming year.

\section*{Acknowledgments} I wish to thank
the Theory Group at Rutgers University,
as well as John Schwarz, Leonard Susskind,
and Ed Witten, who have taught me what little
I know about these subjects.  I particularly
want to thank Tom Banks and Nathan Seiberg for their comments
on the manuscript.  This work supported in part by
the U.S. Department of Energy.


\begin{thebibliography}{99}
\bibitem{apology}
A complete list of references on the subject of string
duality would run many pages.  J. Polchinski~\cite{polchinskicolloquium}\ and
John Schwarz~\cite{schwarzreview}
have
written excellent reviews geared to non-experts
which collect most of the needed references.
Witten has written an article which summarizes
the major ideas.\cite{wittenpt}
As a result, I will only make selected references
in the text, either to
reviews which I believe
may be instructive or to highlight particular points.
\bibitem{apologya}
This particular analogy is an elaboration of one
given by Polchinski in [2].
\end{thebibliography}
\end{document}